\documentclass[9pt,twocolumn,twoside]{opticajnl}
\journal{opticajournal} 

\setboolean{shortarticle}{true}

\usepackage{mathtools}
\usepackage{verbatim}
\usepackage[utf8]{inputenc}
\usepackage[T1]{fontenc}
\usepackage{hyphenat}
\usepackage{graphicx}
\usepackage{amsmath}
\usepackage{geometry}
\usepackage{subcaption}
\usepackage{lineno}
\usepackage{float}

\title{Phase Noise Characterization of Cr:ZnS
Frequency Comb using Subspace Tracking}

\author[1,*]{Aleksandr Razumov}
\author[2]{Sergey Vasilyev}
\author[2]{Mike Mirov}
\author[1]{Jasper Riebesehl}
\author[1]{Holger R. Heebøll}
\author[1]{Francesco Da Ros}
\author[1]{Darko Zibar}

\affil[1]{DTU Electro, Technical University of Denmark (DTU), Kgs. Lyngby, Denmark, 2800}
\affil[2]{IPG Photonics Corporation, 377 Simarano Dr, Marlborough, MA 01752, USA}

\affil[*]{alrazu@dtu.dk}

\begin{abstract}
We present a comprehensive phase noise characterization of a mid-IR Cr:ZnS frequency comb. Despite their emergence as a platform for high-resolution dual-comb spectroscopy, detailed investigations into the phase noise of Cr:ZnS combs have been lacking. To address this, we use a recently proposed phase noise measurement technique that employs multi-heterodyne detection and subspace tracking. This allows for the measurement  of the common mode, repetition-rate and high-order phase noise terms, and their corresponding scaling as a function of a comb-line number, using a single measurement set-up. We demonstrate that the comb under test is dominated by the common mode phase noise, while all the other phase noise terms are below the measurement noise floor ($\sim -120$ dB rad$^2$/Hz), and are thereby not identifiable.    

\end{abstract}

\setboolean{displaycopyright}{false} 

\begin{document}

\maketitle

Over the past decades optical frequency combs became indispensable tools for spectroscopy. For instance, frequency combs greatly enhance capabilities of Fourier transform spectroscopy (FTS) and enable dual comb spectroscopy (DCS) that is simultaneously broadband, precise, high resolution, and fast \cite{Thorpe,Maslowski}. The detailed analysis shows that the performance of comb-based spectroscopy is limited by the optical power of comb modes and by the comb's relative intensity noise (RIN) \cite{Thorpe,Maslowski}. Further, an important new application of frequency combs for the high resolution vacuum ultraviolet (VUV) spectroscopy of the $^{229m}$Th isomeric transition requires the comb sources with exceptionally low phase noise \cite{zhang2024dawnnuclearclockfrequency} . Currently, the VUV combs, which cover 100 nm to 200 nm (1.5 pHz to 3.5 pHz), are generated as high harmonics of available infrared (IR) combs and the power spectral density (PSD) of the comb's phase noise (PN) increases quadratically with harmonic order \cite{Benko2014}.

Recently, ultrafast Cr:ZnS lasers have emerged as a new platform for a high performance FTS and DCS in a broad range of optical frequencies that span from THz to PHz \cite{Krebbers:24,Konnov}. The benefits of Cr:ZnS comb sources encompass high average power (3 W), few-cycle pulses (about 20 fs) and low noise, along with the capability to extend the instantaneous bandwidth through nonlinear up- and down- conversion. Consequently, in the long wave IR regime (6-20 µm, 15-50 THz), the use of Cr:ZnS comb sources resulted in a significant increase of the acquisition speed of DCS: about quarter million of comb modes were resolved in real time \cite{Konnov}, while DCS with established platforms has required  minutes of averaging to obtain similar comb-mode resolved spectra  \cite{Abijith}. In the UV regime, Cr:ZnS-based combs enabled the first broadband DCS: a million of comb modes were resolved in the 325-342 nm (876-922 THz) band, versus 130 resolved modes reported in \cite{Xu24}.

There have been a number of recent reports that discussed the intensity noise of ultrafast Cr:ZnS lasers and amplifiers and nonlinear noise suppression via $\chi^{(2)}$ process in polycrystalline gain elements \cite{Vasilyev2024, Bu:22}. However, to the best of our knowledge a comprehensive phase noise characterization of Cr:ZnS-based combs has been missing. 

In general, performing phase noise characterization of frequency combs is challenging as many of the combs parameters can exhibit random fluctuations, i.e.~carrier, envelope offset frequency, repetition rate, pulse duration, pulse shape and chirp \cite{Paschotta2006}. The majority of the state-of-the-art frequency comb phase noise characterization techniques rely on line-by-line measurements \cite{Kim16}, which are tedious and do not give a comprehensive picture. For instance, a delayed self-heterodyne method (DSH), which is the most popular technique, requires a precise optical filtering of the comb-line of the interest \cite{Roos:04}. This becomes a severe problem when we want to characterize frequency combs with the free spectral range (FSR) of several tens of MHz and below. Moreover, DSH relies on fiber delays that are much longer than the coherence length of individual lines. They are prone to temperature fluctuation and require active stabilization. Short delay lines can be employed, but these require careful calibration and de-convolution techniques \cite{Yuan:22}.


\begin{figure*}[ht]
    \centering
    \includegraphics[width=\textwidth]{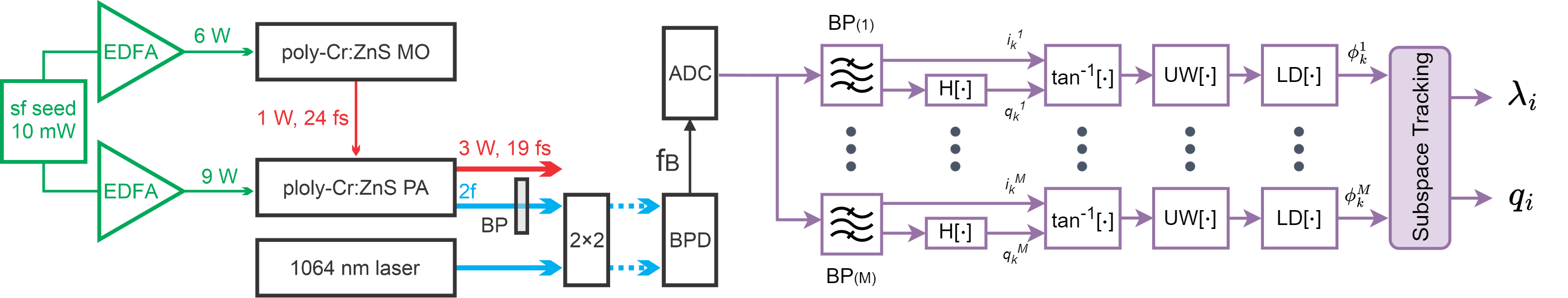}
    \caption{Experimental setup . The $2f$ output of the Cr:ZnS comb is band-pass filtered (BP) and superimposed with a radiation from a narrowband 1.064 $\mu m$ laser in a 2×2 fiber coupler. ADC - Analog-to-Digital Converter}
    \label{fig:setup}
\end{figure*}

Most importantly, getting the full picture of frequency comb phase noise dynamics, requires identifying and measuring various phase noise contributions that add up to the total phase noise, $\phi_m(t)$, of the $m$-th comb-line: $\phi_m(t) = \phi^{cm}(t) + m\phi^{rep}(t) + \phi^{res}(m,t)$. $\phi^{cm}(t)$ represents the common mode phase noise (common to all lines), and $\phi^{rep}(t)$ is the repetition rate phase noise, that contributes to the timing jitter of the pulses, and scales linearly with comb-line number $m$ \cite{Benkler}. Finally, fluctuations stemming from internal or external noise sources, and which do not have a straightforward mathematical description nor scaling with comb-line number, can also contribute to the overall phase noise. We lump all this contributions in a term $\phi^{res}(m,t)$ \cite{Benkler,Paschotta2006,Razumov2023}.

We have recently proposed a measurement technique based on multi-heterodyne detection and subspace tracking able to identify and measure various phase noise contributions associated with the $m$-th comb-line total phase noise $\phi_m(t)$.
The proposed measurement technique is able to decompose total phase noise, $\phi_m(t)$, in terms of $\phi^{cm}(t)$, $\phi^{rep}(t)$ and $\phi^{res}(m,t)$, and provide their corresponding power spectrum densities and scaling with comb-line number \cite{Razumov2023}. In \cite{RazumovCLEO24}, we employ the aforementioned technique, and perform comprehensive phase noise characterization of the mid-IR frequency comb based on Cr:ZnS modelocked laser. In this Letter, we expand upon our prior work presented in \cite{RazumovCLEO24} by extending the measurement range and improving on the signal processing methods. More precisely, our investigation now includes a broader range of spectral lines (25 compared to the original 5). Moreover, we have improved the phase detrending algorithm, which has resolved an ambiguity related to the repetition rate phase noise, i.e.~$\phi^{rep}(t)$. The experimental set-up including the digital signal processing block for the phase noise extraction and analysis is shown in Fig.~\ref{fig:setup}.

The design of the mid-IR ($\lambda_c=2.4 \mu m$) Cr:ZnS frequency comb is described in detail in \cite{Vasilyev2019}. In short, the comb consists of a Cr:ZnS master oscillator (MO) ($1 W, 24 fs$) that is coupled to a single-pass Cr:ZnS power amplifier (PA) at full repetition rate ($f_R=80$ MHz). Both components are optically pumped by low-noise Er-doped fiber amplifiers (EDFA) seeded by a single frequency (sf) semiconductor laser (Fig. \ref{fig:setup}). The polycrystalline Cr:ZnS amplifier features simultaneous amplification of pulses and generation of optical harmonics that occurs due to the random quasi phase matching process. The second harmonic signal ($2f$) is used to measure heterodyne beatings between the spectral components of the comb and an ultra-narrowband laser provided by TimeBase ($1.064 \mu m$, linewidth $< 4$ Hz in 1 second), that acts as the Local Oscillator (LO). The LO laser is stabilized to ultra-low expansion glass cavity, and is expected to have significantly lower-phase noise than the comb.  

\begin{figure}[h]
    \centering
    \includegraphics[width=0.9\linewidth]{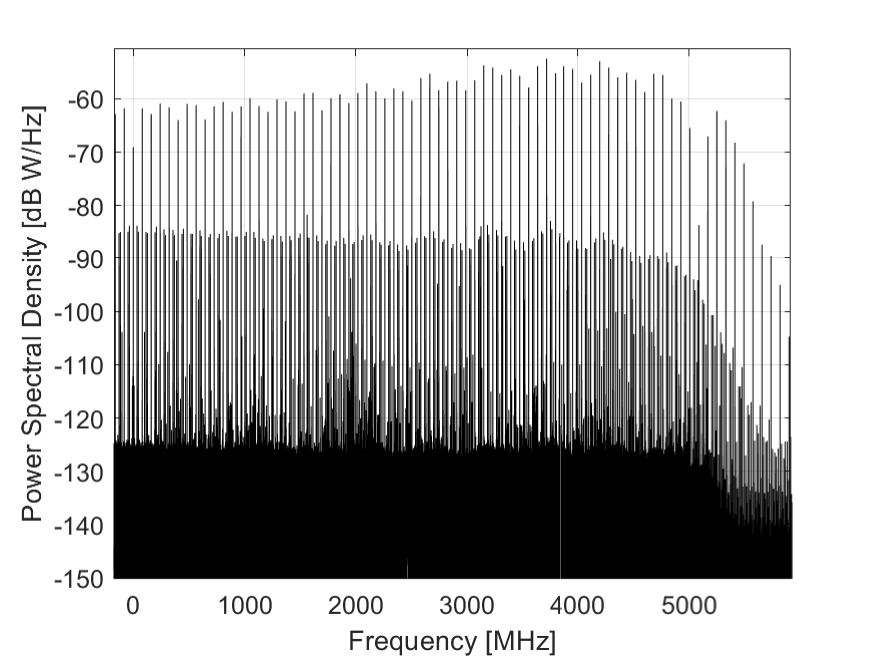}
    \caption{Spectrum of the downconverted Cr:ZnS comb signal}
    \label{fig:spectr}
\end{figure}

A balanced receiver with 4 GHz bandwidth followed by digital sampling scope ($Fs=50$ GSa/s, and $BW=10$ GHz) is employed for the measurement of the beat heterodyne signal. We assume that the phase noise of the beat heterodyne signal is dominated by the phase noise originating from the frequency comb. Digital signal processing, including phase estimation and detrending as well as subspace tracking, described in \cite{Razumov2023}, are then performed in offline mode. Bandwidth of the optical bandpass filter allows us to capture around 125 comb-lines with sufficient signal-to-noise ratio, see downconverted spectra in Fig. \ref{fig:spectr}. Because of the aliasing the spacing between the downconverted comb-lines in Fig.~\ref{fig:spectr} is around $40$ MHz. Only 25 comb-lines are used for signal processing due to the random access memory (RAM) constraints.  

In the following, for the consistency of the paper, we briefly summarize the main idea and concept behind the subspace tracking. A detailed description can be found in \cite{Razumov2023}.  

We assume that for the comb under consideration, we have $P=3$ phase noise sources, i.e. common mode, repetition rate and residual phase noise, i.e.~$\boldsymbol{\phi}^s(t) = [\phi^{cm}(t),\phi^{rep}(t),\phi^{res}(t)]^T$, respectively. After signal detection, we extract phase noise of each of $M$ detected comb-lines, $\boldsymbol{\phi}(t)=[\phi^{-(M-1)/2}(t),...,\phi^{(M-1)/2}(t)]^T$, where $T$ is a transpose operator. Each of the phase noise components of $\boldsymbol{\phi}(t)$, contain the total phase noise which is a linear combination of phase noise originating form the common-mode, repetition rate and residual phase noise. In matrix form, this is expressed as:
     $ \boldsymbol{\phi}_m(t) =\mathbf{H}\boldsymbol{\phi}^s(t)$

The matrix $\mathbf{H}$ is a (generation) matrix of $M\times P$ dimension containing information how each phase noise source $[\phi^{cm}(t),\phi^{rep}(t),\phi^{res}(t)]^T$ contributes to the phase noise of each of the detected comb lines of $\boldsymbol{\phi}(t)$. It is expressed as:
\begin{equation}
    \mathbf{H} = \begin{bmatrix}
1 & -(M-1)/2 & h_{13}\\
\vdots & \vdots & \vdots\\
1 & (M-1)/2 & h_{M3}
\end{bmatrix}
\end{equation}

The first and the second column of the matrix $\mathbf{H}$ indicate that the common mode and the repetition rate phase noise terms, $\phi^{cm}(t)$ and $\phi^{rep}(t)$, contribute equally and linearly, respectively, to the total phase noise of the detected comb lines. Since we don't know how the residual phase noise term contributes to the total phase noise, we denote its contribution by coefficients, $[h_{13},...h_{M3}]^T$.   

The objective of the subspace tracking is to estimate phase noise components associated with $\boldsymbol{\phi}^s(t)$. Once $\boldsymbol{\phi}^s(t)$ is determined, power spectrum density (PSD) of the common mode, repetition rate and residual phase noise can be computed. Recovering  $\boldsymbol{\phi}^s(t)$ is achieved by finding a projection matrix $\mathbf{G}$ with dimensions $P \times M$ such that: $\boldsymbol{\phi}
    ^s(t) = \mathbf{G} \boldsymbol{\phi}_m(t)$

It has been shown in \cite{Razumov2023} that the matrix $\mathbf{G}$ can be found by setting $\mathbf{G} = \mathbf{Q^T_P}$, where matrix $\mathbf{Q_P}$ contains $P$ eigenvectors of the phase noise covariance matrix $\mathbf{S}^{M\times M}$ of the detected phase noise components $\boldsymbol{\phi}(t)$. We will now explain how to determine the phase noise correlation matrix for the experimental set-up in Fig.~\ref{fig:setup}.  

Prior to subspace tracking we acquire phase noise traces $\phi^k_m$ of the measured frequency comb-lines, where $m=1,...,M$ and $k=1,...,K$ is a discrete time index as the output of the balanced detector is sampled. $M$ and $K$ represent the number of detected comb lines and the length of the discrete time-domain phase noise trace, respectively. 
This is achieved using the phase estimation method which utilizes a bank of bandpass (BP) filters with a bandwidth of $B=24$ MHz, $BP_{(1)}$ to $BP_{(M)}$ in Fig.~\ref{fig:setup}. These filters extract different frequency comb lines from the detected beat signal. For each filtered line $i_k^m$, a Hilbert transform is applied to recover the corresponding orthogonal quadrature $q_k^m$. The phase of the $m$-th comb line is then determined by taking the inverse tangent, denoted as $\tan^{-1}(i_k^m/q_k^m)$, of the reconstructed field. To ensure smoothness, the resulting phases undergo an unwrapping function that corrects large phase jumps exceeding $\pi$ by adding multiples of $2\pi$, thereby producing a continuous phase function over time. Finally, a linear detrending process is applied to the phase traces yielding phase noise estimates $\phi^k_m$ for each of the $M$ comb lines. Obtained phase noise traces $\phi_m^k$ are then employed to construct the sample covariance matrix, compute eigenvectors and eigenvalues, and estimate source of phase noise, $\boldsymbol{\phi}^s(t)$, associated with a Cr:ZnS comb.

The next step is computing the phase noise sample covariance matrix $S(K)$ as follows:
\begin{equation}
    S(K) = \frac{1}{K - 1}\sum^K_{k=1}\phi_k\phi_k^T
\end{equation}
The matrix $S(K)$ can then be factored into: $S = Q\Lambda Q^T$, where $\Lambda$ is a matrix containing eigenvalues and $Q$ is a matrix containing eigenvectors. 

The outcome of the subspace tracking in terms of the eigenvalues and the eigenvectors can be related to the physical properties of an optical frequency comb \cite{Razumov2023}.  
The eigenvalues are proportional to the variance of phase noise sources, i.e. $\phi^{cm}(t), \phi^{rep}(t)$ and $\phi^{res}(m,t)$. The eigenvectors indicate how different phase noise sources contribute to various comb-lines. Finally, by projecting $\boldsymbol{\phi}_m(t)$ onto eigenvectors we can obtain $\phi^{cm}(t)$, $\phi^{rep}(t)$, and $\phi^{res}(m,t)$, and the corresponding power spectrum densities can be computed then.

\begin{figure}[h!]
  \centering
    \begin{subfigure}{\linewidth}
        \includegraphics[width=0.9\linewidth]{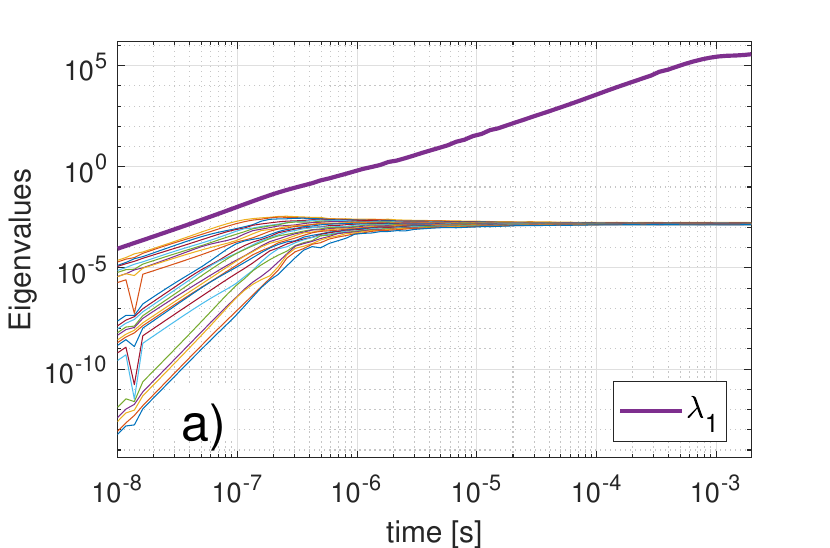}
    \end{subfigure}
    \begin{subfigure}{\linewidth}   \includegraphics[width=0.9\linewidth]{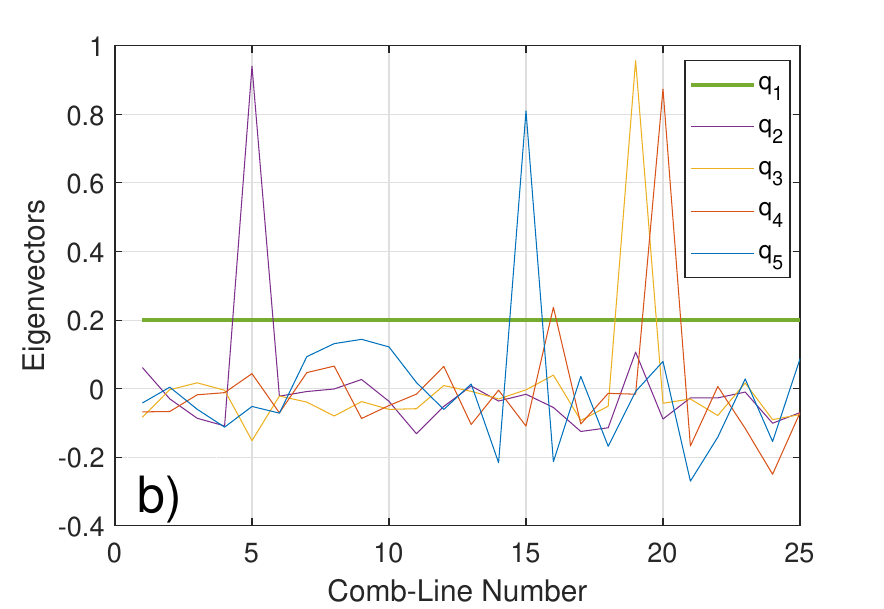}
    \end{subfigure}	
    \begin{subfigure}{\linewidth}   \includegraphics[width=0.9\linewidth]{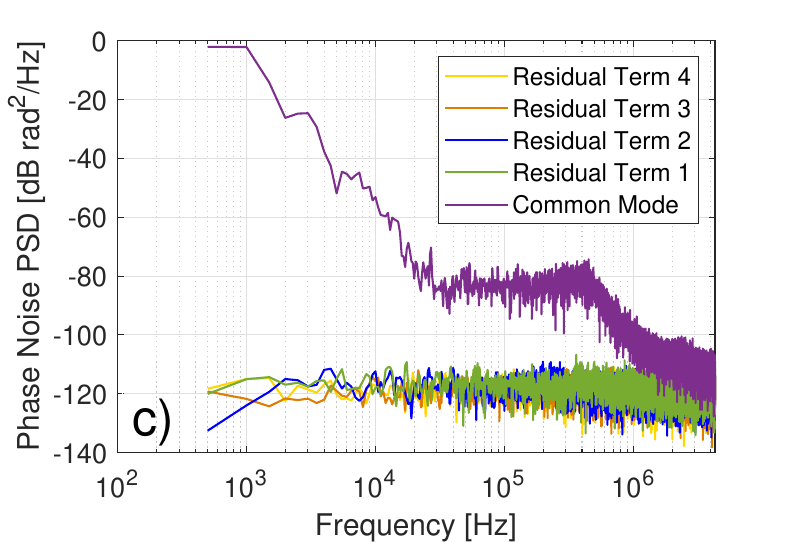}
    \end{subfigure}	
\caption{The result of subspace tracking: (a) Evolution of eigenvalues of sample covariance matrix $S(K)$, as a function of time t; (b) Evolution of eigenvectors of sample covariance matrix $S(K)$, as a function of comb-line number $m$; (c) Power Spectral Density of the phase noise sources, as a function of frequency}
\label{plots}
\end{figure}

Fig. \ref{plots} (a) illustrates the evolution of eigenvalues as a function of time. We can clearly observe only one eigenvalue that increases with time, indicating that there is only one detectable source of phase noise. Eigenvalues that are constant indicate a measurement noise floor. We observe that the first eigenvalue, $\lambda_1$, is clearly dominating.  

In Fig.~\ref{plots}(b), the eigenvectors $[\mathbf{q}_1,...\mathbf{q}_5]$, associated with the phase noise covariance matrix are shown. The first eigenvector, $\mathbf{q}_1$, associated with $\lambda_1$, is constant as a function of comb-line number. This implies that the phase noise source associated with $\mathbf{q}_1$, and obtained by projecting $\boldsymbol{\phi}^k_m$ on $\mathbf{q}_1$, will contribute equally to all comb-lines under consideration, i.e. common mode phase noise source, $\phi^{cm}(t)$.

According to the elastic tape model, it is expected that the second eigenvector $\mathbf{q}_2$  scales linearly as a function of comb-line number. By projecting $\boldsymbol{\phi}^k_m$ on $\mathbf{q}_2$, we would then able to obtain repetition rate phase noise, i.e~$\phi^{rep}(t)$. However, looking at Fig.~\ref{plots}(b) expected linearity in $\mathbf{q}_2$ is not observed. Instead, random fluctuations are apparent.

The reason for this is that the repetition rate phase noise $\phi^{rep}(t)$ is too low to be measured, i.e.~it is hidden below the measurement noise floor (-120 dB rad$^2$/Hz). From a physical perspective it means that we are not able to estimate a pulse timing jitter with a given amount of measurement noise in the system. 
Therefore, the remaining eigenvectors $[\mathbf{q}_2-\mathbf{q}_5]$, corresponding to $\lambda_2-\lambda_5$, are associated with the measurement noise floor. Only 5 eigenvectors $[\mathbf{q}_1-\mathbf{q}_5]$ are shown in order to keep the figure readable. Other eigenvectors $[\mathbf{q}_6-\mathbf{q}_{25}]$  exhibit similar to $[\mathbf{q}_2-\mathbf{q}_5]$  behaviour and do not contain useful information.

Fig. \ref{plots} (c) presents the phase noise PSDs associated with different eigenvectors, i.e. common mode, repetition rate, residual phase noise terms and other. 

It is observed that the common mode phase noise associated with the the first eigenvector and eigenvalue dominates. The phase noise PSD associated with the second eigenvalue and eigenvector, repetition rate phase noise, (Residual term 1), is constant and thereby dominated by the measurement noise floor. The remaining PSDs associated with the third, fourth and fifth eigenvector are likewise dominated by the measurement noise floor, Residual Term 2 to 4. Only first five phase noise source PSDs were shown for the sake of readability. For the given experiment the measurement noise floor is observed to be at approximately -120 dB $\mathrm{rad^2}$/Hz.

\begin{figure}[h]
    \centering
    \includegraphics[width=0.87\linewidth]{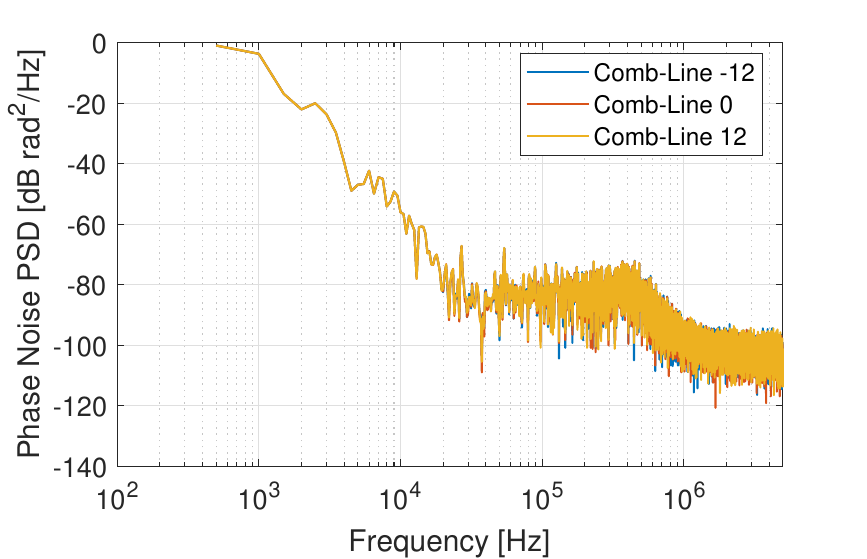}
    \caption{Phase Noise Power Spectral Density $\phi_m$ of comb-lines $m = -12, 0, 12$, as a function of frequency}
    \label{fig:psd3}
\end{figure}

It is noteworthy that due to the measurement of the second harmonic signal ($2f$), the  phase noise is multiplied by a factor two, which translates into 6 dB higher PSD. \cite{Benko2014}.

To highlight the fact that the timing jitter contribution to the total comb-line phase noise is low, we show phase noise power spectral densities of the most distant comb-lines $m=-12, 0, 12$, see Fig. \ref{fig:psd3}. As can be seen from Fig. \ref{fig:psd3} all PSDs plots overlap, indicating insignificance of $\phi^{rep}(t)$.

\begin{figure}[h]
    \centering
    \includegraphics[width=0.87\linewidth]{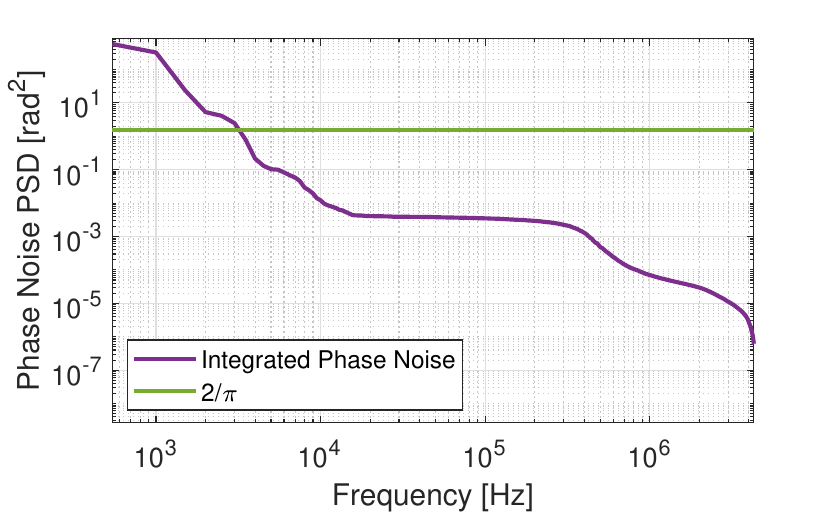}
    \caption{Integrated common mode phase noise
    }
    \label{fig:integral}
\end{figure}

\begin{equation}
\label{lw}
    \int_{\frac{\delta\nu}{2}}^{+\infty} S_{\phi}(f)df = \frac{2}{\pi}
\end{equation}

To estimate Full Width at Half Maximum (FWHM) linewidth associated withe the common mode phase noise term, we have employed Eq.~\ref{lw}, where $\delta\nu$ is a FWHM linewidth, $S_{\phi}(f)$ is the common mode phase noise PSD, and $f$ is the frequency. Fig. \ref{fig:integral} shows integrated common mode phase noise. As illustrated in the figure, the integrated phase noise reaches $2/\pi$ approximately at 3.6 kHz. Consequently, the FWHM linewidth can be approximated to be ~7.2 kHz.

We performed, a comprehensive characterization of the phase noise in the Cr:ZnS frequency comb. The measurements show that the phase noise of the comb is completely dominated by the common mode phase noise, within the given dynamic range of the measurement imposed by the measurement noise floor. The repetition rate and the higher order phase noise terms were not observed, and must thereby be below the measurement noise floor of $-120$ dB rad$^2$/Hz.


\begin{backmatter}
\bmsection{Funding} Villum Fonden VI-POPCOM VIL54486 \& OPTIC-AI
VIL29334

\bmsection{Acknowledgments} We thank Sujit Basu, William Dixon, Alan Wolke, all with Tektronix, for equipment loans and consulting on signal acquisition.

\bmsection{Disclosures} The authors declare no conflicts of interest.

\bmsection{Data availability} Data underlying the results presented in this paper are not publicly available at this time but may be obtained from the authors upon reasonable request.

\end{backmatter}

\bibliography{Optics_Letters}



\ifthenelse{\equal{\journalref}{aop}}{%
\section*{Author Biographies}
\begingroup
\setlength\intextsep{0pt}
\begin{minipage}[t][6.3cm][t]{1.0\textwidth} 
  \begin{wrapfigure}{L}{0.25\textwidth}
    \includegraphics[width=0.25\textwidth]{john_smith.pdf}
  \end{wrapfigure}
  \noindent
  {\bfseries John Smith} received his BSc (Mathematics) in 2000 from The University of Maryland. His research interests include lasers and optics.
\end{minipage}
\begin{minipage}{1.0\textwidth}
  \begin{wrapfigure}{L}{0.25\textwidth}
    \includegraphics[width=0.25\textwidth]{alice_smith.pdf}
  \end{wrapfigure}
  \noindent
  {\bfseries Alice Smith} also received her BSc (Mathematics) in 2000 from The University of Maryland. Her research interests also include lasers and optics.
\end{minipage}
\endgroup
}{}

\end{document}